\documentclass[conference]{IEEEtran}
\IEEEoverridecommandlockouts
\usepackage{cite}
\usepackage{amsmath,amssymb,amsfonts}
\usepackage{graphicx}
\usepackage{textcomp}
\usepackage{xcolor}
\usepackage{subcaption}
\usepackage{float}
\usepackage{booktabs} % 提供 \toprule, \midrule, \bottomrule
\usepackage{multirow}
\usepackage{amsmath}
\usepackage{algorithm}
\usepackage{algpseudocode}
\usepackage{fancyhdr}
\def\BibTeX{{\rm B\kern-.05em{\sc i\kern-.025em b}\kern-.08em
    T\kern-.1667em\lower.7ex\hbox{E}\kern-.125emX}}
\IEEEoverridecommandlockouts
\usepackage[letterpaper,top=0.73in,bottom=1.09in,left=0.63in,right=0.63in]{geometry}

\fancypagestyle{IEEEtitlepagestyle}{%
    \fancyhf{}
    \fancyhead[L]{\normalsize Y. Guo, T. Zhang, X. Gu, S. Sun, M. Tao, and R. Gao, ``Directional Measurements and Analysis for FR3 Low-Altitude Channels in a Campus Environment,'' \textit{IEEE/CIC International Conference on Communication in China (ICCC Workshops)}, Shanghai, China, 2025.}
    
}
\fancypagestyle{plain}{%
    \fancyhf{}

}

\begin{document}

% 强制使用页眉样式
\pagestyle{fancy}
\thispagestyle{IEEEtitlepagestyle}

\title{Directional Measurements and Analysis for FR3 Low-Altitude Channels in a Campus Environment\\
\thanks{This work is supported in part by the National Natural Science Foundation of China under Grants 62431014, 62271310, 62125108, and 62001254, and in part by the Natural Science Foundation of Nantong under Grant JC2023074.}
}
\setlength{\columnsep}{0.241in}
\author{
    \IEEEauthorblockN{Yulu Guo$^{1}$, Tongjia Zhang$^{1}$, Xiangwen Gu$^{1}$, Shu Sun$^1$, Meixia Tao$^1$, Ruifeng Gao$^{2, 3}$}
    \IEEEauthorblockA{$^1$ School of Information Science and Electronic Engineering, Shanghai Jiao Tong University, Shanghai 200240, China}
    \IEEEauthorblockA{$^2$ School of Transportation and Civil Engineering, Nantong University, Nantong 226019, China}
    \IEEEauthorblockA{$^3$ Nantong Research Institute for Advanced Communication Technologies, Nantong 226019, China}
    \IEEEauthorblockA{Corresponding author: Shu Sun (e-mail: shusun@sjtu.edu.cn)}
 }
\maketitle
\begin{abstract}
In this paper, we present detailed low-altitude channel measurements at the FR3 band in an outdoor campus environment. Using a time-domain channel sounder system, we conduct two types of measurements: path loss measurements by moving the transmitter (Tx) at one-meter intervals along a 26-point rooftop path, and directional power angular spectrum measurements through antenna scanning at half-power beam width intervals. The path loss analysis across different Rx shows that the close-in model outperforms conventional 3GPP models and height-corrected variants, with path loss exponents close to free space values indicating line-of-sight dominance. 
The power angular spectrum measurements show that propagation behavior varies significantly with environmental conditions. Closer Rx exhibit stronger sensitivity to ground reflections during downward Tx tilting, while obstructed links display uniform angular characteristics due to dominant scattering effects, and corridor environments produce asymmetric power distributions.
These results indicate that low-altitude propagation is characterized by complex interactions between Tx height and ground scattering mechanisms, providing fundamental insights for channel modeling in emerging mid-band communication systems.
\end{abstract}

\begin{IEEEkeywords}
Channel measurement, low-altitude Channel, path loss, power-angular characteristics, mid-band frequency.
\end{IEEEkeywords}

\section{Introduction}
The evolution toward sixth-generation (6G) wireless systems has intensified the exploration of new spectrum bands to meet unprecedented demands for ubiquitous connectivity and spectral efficiency. The FR3 mid-band frequency range (7-24 GHz) has been seen as a “golden band" for 6G applications \cite{Shakya24OJCS}, offering an optimal balance between propagation characteristics and system capacity. Concurrently, the rise of low-altitude platforms, including unmanned aerial vehicles (UAVs), aerial base stations, and urban air mobility systems, calls for a comprehensive understanding of radio wave propagation in three-dimensional environments at these emerging frequencies\cite{Khawaja2019CST}.

Channel measurement is the most accurate way to obtain realistic channel characteristics, especially for emerging frequency bands \cite{Shakya24OJCS, Rappaport13Access, Mao24TIM}, and recent years have witnessed a series of measurement campaigns targeting low-altitude propagation across various frequencies. Wang et al. \cite{Wang17VTC} conducted air-to-ground channel measurements at 2.412 GHz and 919 MHz using UAVs flying at 25-150 m altitudes, proposing modified path loss models with height-dependent correction factors for cellular-connected UAV applications. Cui et al. \cite{Cui20Access} carried out comprehensive channel characterization at 1, 4, 12, and 24 GHz in suburban scenarios, analyzing path loss exponents (PLEs), shadow fading, and small-scale fading statistics to demonstrate frequency-dependent propagation behaviors. For specialized applications, Li et al. \cite{Li23Drone} developed a machine-learning-based path loss model for agricultural UAV communications through 3.6 GHz farmland measurements, achieving superior prediction accuracy compared to traditional empirical models. At millimeter-wave frequencies, Semkin et al. \cite{Semkin21PIMRC} developed a lightweight 1.3 kg channel measurement system at 28 GHz, conducting airborne measurements over 300 m distances while addressing UAV frame effects on antenna patterns through calibration procedures.

Since current measurement studies specifically targeting low-altitude scenarios in the FR3 frequency range are limited, our work provides valuable insights by conducting both path loss and power angular spectrum measurements using a time-domain channel sounder system. The path loss analysis demonstrates that the close-in (CI) free space reference distance model outperforms conventional 3GPP models and height-corrected variants, while the angular spectrum characterization reveals distinct environment-dependent propagation behaviors, including ground reflection effects and asymmetric power distributions in constrained environments. 

\section{Measurement Campaign}
\subsection{Measurement System}
In this work, we employed a time-domain channel sounder system based on National Instruments hardware to conduct the channel directional measurement campaign. The channel sounder utilizes field-programmable gate arrays and operates in a superheterodyne architecture with a flexible intermediate frequency range of 8 to 12 GHz. The system transmits a Zadoff-Chu sequence with a length of 65,535 at a center frequency of 10 GHz with a transmit power of 10 dBm \cite{Guo25WCNC}. Both the transmitter (Tx) and receiver (Rx) are equipped with identical horn antennas, each providing 20 dBi gain and a half-power beam width of 30° in both azimuth and elevation planes. Channel impulse responses (CIRs) and power delay profiles (PDPs) are obtained through correlation processing at the Rx, achieving a maximum delay resolution of 650 ps. Synchronization between the transmitter and Rx is maintained using two pre-synchronized rubidium clocks. Detailed measurement system parameters are listed in Table \ref{tab:system_parameters}.

\subsection{Measurement Scenario}
\begin{figure}[!t]
    \centering
    \includegraphics[width=1\linewidth]{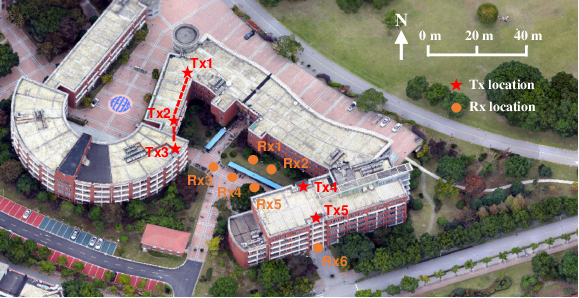}
    \caption{Map of the measurement scenario showing Tx and Rx locations.}
    \label{fig:measurement_setting}
\end{figure}

Our directional measurements were conducted on the Minhang campus of Shanghai Jiao Tong University. A total of 4 TX and 6 Rx locations were selected, covering rooftops of college buildings as well as open spaces between buildings, as shown in Fig. \ref{fig:measurement_setting}. Three Tx (Tx1, Tx2, and Tx4) were positioned on rooftops at heights of approximately 25-30 m to simulate low-altitude UAV flight scenarios. Tx3 was placed on the ground directly beneath Tx2 to provide a reference ground-level link. All Rx were positioned at ground level with antenna heights of 1.5 m and distributed across open spaces and passages within the central campus area. This configuration enabled comprehensive characterization of various propagation scenarios, including line-of-sight (LoS) and non-line-of-sight (NLoS) propagation paths. For each measurement position, we ensured that the Tx antenna and Rx antenna were in the far field of each other \cite{Sun25CM}. 

\begin{table}[!t]
\caption{\textsc{Measurement System Parameters}}
\centering
\begin{tabular}{|c|c|}
\hline
\textbf{Parameter} & \textbf{Value} \\
\hline
Carrier Frequency & 10 GHz \\
\hline
Bandwidth & 2 GHz \\
\hline
Tx Antenna Height & 1.5 m, 25 m, 30 m \\
\hline
Rx Antenna Height & 1.5 m \\
\hline
Transmit Power & 10 dBm \\
\hline
\multirow{3}{*}{Antenna Type} & Horn Antenna \\
\cline{2-2}
& HPBW: 30° (Azimuth and Elevation) \\
\cline{2-2}
& Antenna Gain: 20 dBi \\
\hline
System Power Amplifier & 10.01 dB \\
\hline
Baseband Signal & ZC Sequence (length = 16 order) \\
\hline
Multipath Delay Resolution & 650 ps \\
\hline
\end{tabular}
\label{tab:system_parameters}
\end{table}

\begin{figure}[!t]
    \centering
    \includegraphics[width=1\columnwidth]{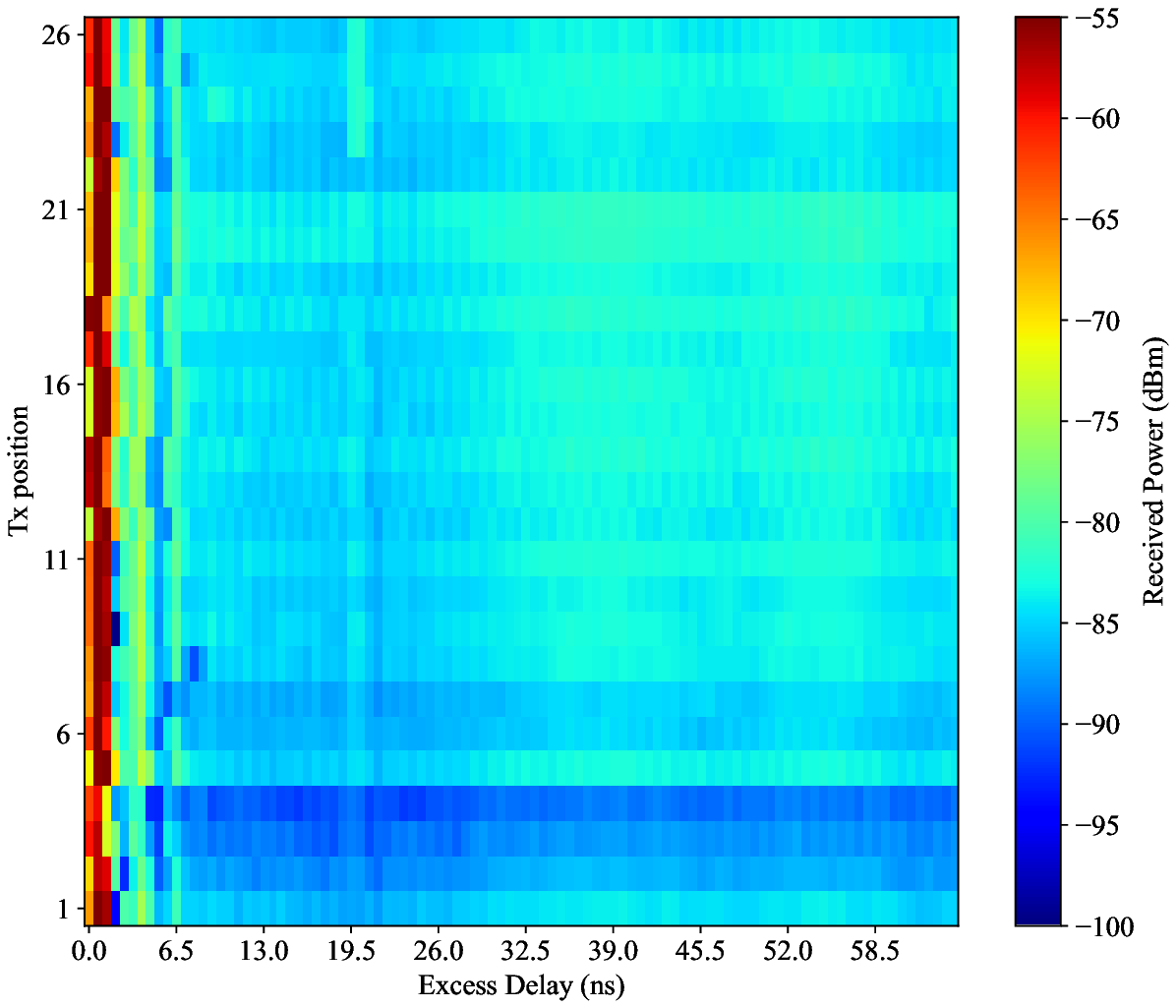}
    \caption{An example of PDPs received by Rx1 vs Tx position.}
    \label{fig:PDP received by Rx1 vs Tx position}
\end{figure}

\subsection{Measurement Procedure}
Due to the large size, high weight, and power consumption of the time-domain channel sounder system used in this measurement, which features real-time time-domain signal processing capability and wide bandwidth, it is currently impractical to mount the system on small-to-medium-sized UAVs for airborne measurements. Therefore, we conducted sampling measurements along the line connecting Tx1 (starting point) and Tx2 (endpoint) by moving the Tx at one-meter intervals. A total of 26 sampling points were measured to simulate the flight trajectory of a UAV at this altitude. An example of PDPs received by Rx1 is shown in Fig. \ref{fig:PDP received by Rx1 vs Tx position}. Throughout the movement, LoS alignment between the Tx and Rx antennas was maintained. The Rx were positioned at ground-level locations Rx1 and Rx2, with some vegetation obstruction between them.

For the measurements involving Tx-Rx combinations of Tx2/Tx3/Tx4 with Rx3/Rx4/Rx5/Rx6, we conducted power angular spectrum (PAS) measurements by varying both horizontal and vertical antenna pointing angles at the Tx and Rx sides after establishing initial antenna alignment. 
The angular measurement procedure follows the double directional channel characterization framework \cite{Shakya25WCNC, Ju21JSAC}, where the CIR $h_{omni}(t, \vec{\Theta}, \vec{\Phi})$ is characterized by azimuth angle of departure (AOD), zenith angle of departure (ZOD), azimuth angle of arrival (AOA), and zenith angle of arrival (ZOA):

\begin{align}
h_{omni}(t, \vec{\Theta}, \vec{\Phi}) &= \sum_{n=1}^{N} \sum_{m=1}^{M_n} a_{m,n} e^{j \varphi_{m,n}} \cdot \delta(t - \tau_{m,n}) \nonumber \\
&\quad \cdot \delta(\vec{\Theta} - \vec{\Theta}_{m,n}) \cdot \delta(\vec{\Phi} - \vec{\Phi}_{m,n}),
\label{eq:channel_model}
\end{align}

\noindent where $\vec{\Theta} = (\phi_{AOD}, \theta_{ZOD})$ represents the 3D TX pointing direction vector, and $\vec{\Phi} = (\phi_{AOA}, \theta_{ZOA})$ is the 3D RX pointing direction vector. The parameters $N$, $M_n$, $a_{m,n}$, $\varphi_{m,n}$, and $\tau_{m,n}$ represent the number of clusters, cluster subpaths, subpath magnitude, phase, and propagation delay, respectively. The PAS measurements aim to characterize the power distribution across these directional vectors at both the Tx and Rx sides.

The angular scanning was performed sequentially for the Tx and Rx sides. First, with the TX antenna pointing toward the RX, the RX antenna orientation was systematically varied across different azimuth and elevation angles at 30° intervals to characterize the AOA and ZOA PAS. Subsequently, the RX antenna was oriented toward the TX, and the TX antenna pointing direction was varied across different AOD and ZOD angles at 30° intervals while recording the received power. 

\section{Measurement Results, Modeling, and Analysis}

\subsection{Path Loss Modeling}

\begin{figure}[!t]
    \centering
    \begin{subfigure}[t]{\columnwidth}
        \centering
        \includegraphics[width=1\columnwidth]{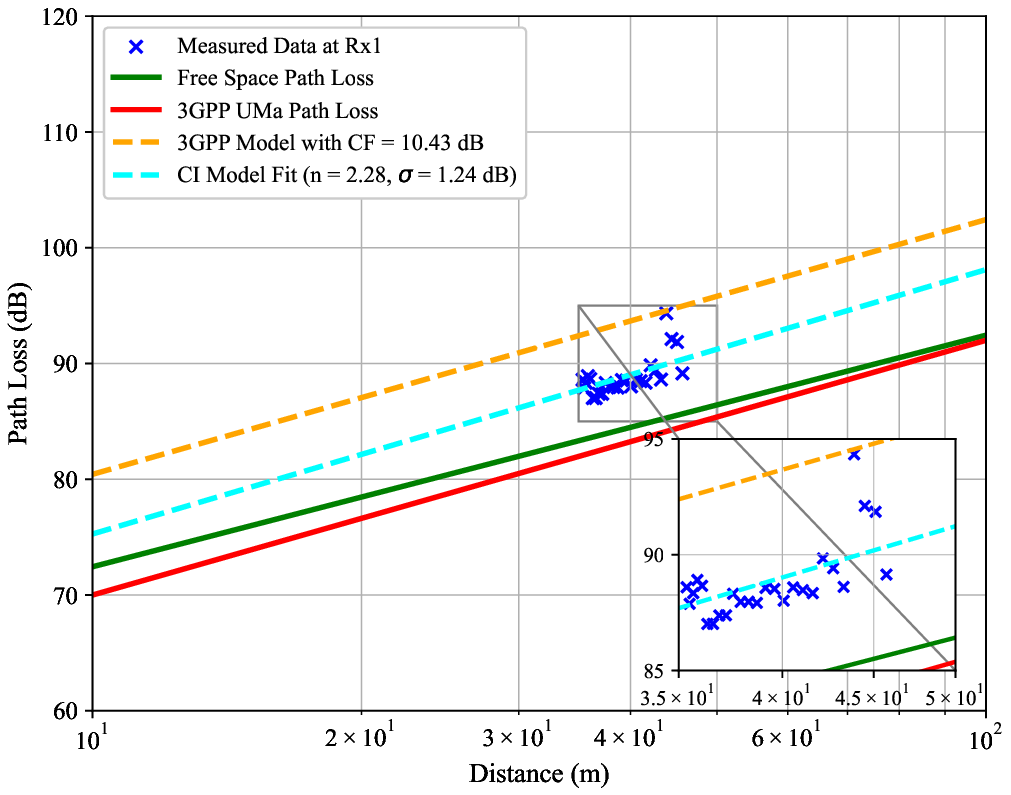}
        \caption{Rx1}        
    \end{subfigure}
    \begin{subfigure}[t]{\columnwidth}
        \centering
        \includegraphics[width=1\columnwidth]{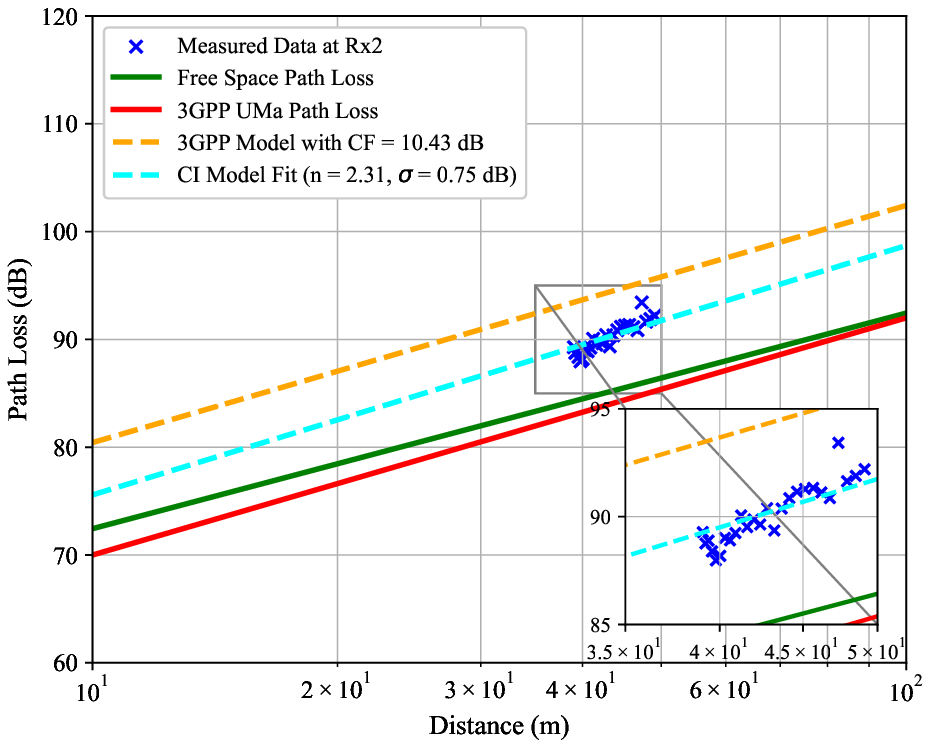}
        \caption{Rx2}       
    \end{subfigure}   
    \caption{Path loss measurement and modeling for different Rx positions.}
    \label{fig2}
\end{figure}

\begin{figure*}[!t]
    \centering
    \begin{subfigure}[t]{0.49\textwidth}
        \centering
        \includegraphics[width=1\textwidth]{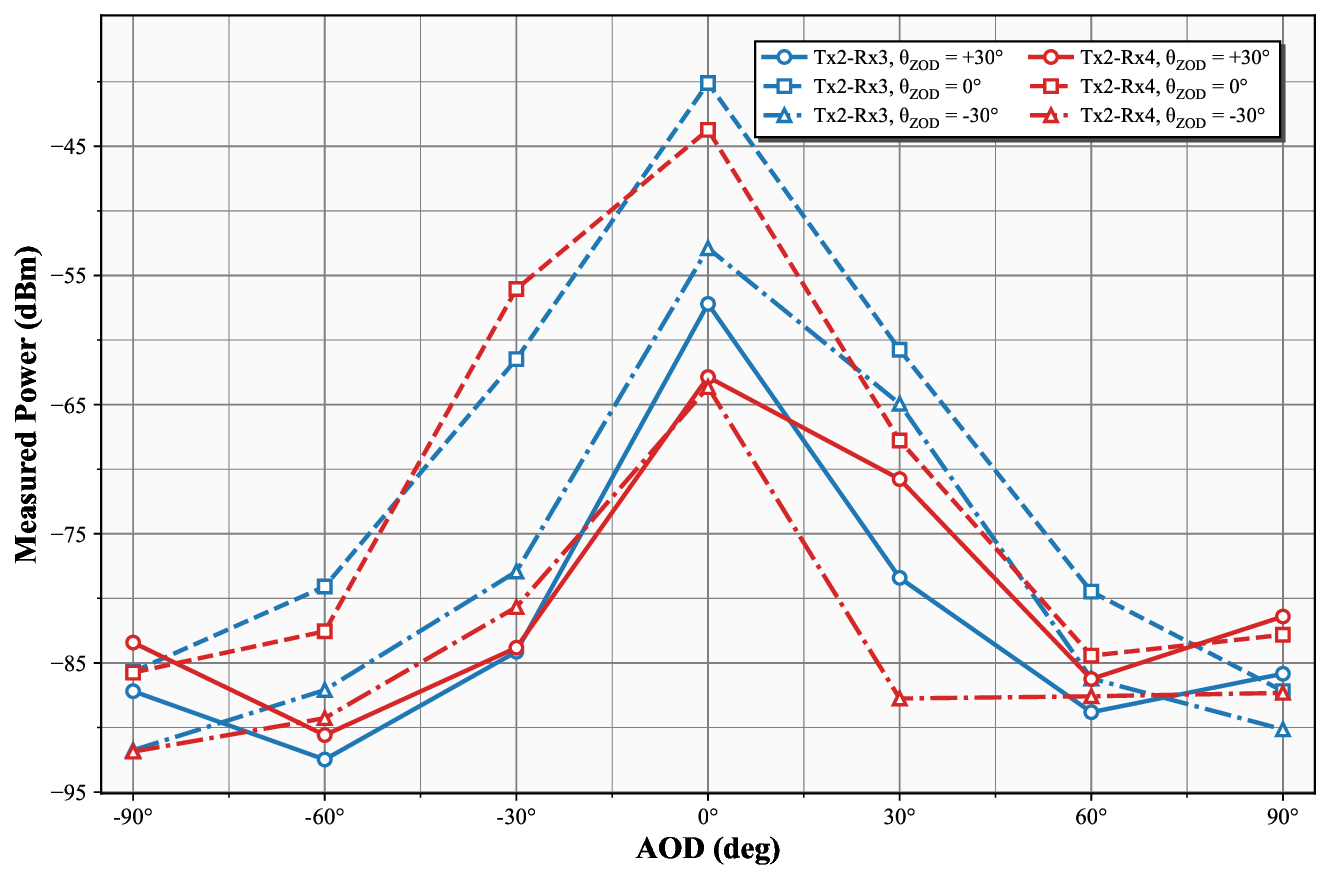}
        \caption{Tx2-Rx3/Rx4}        
    \end{subfigure}
    \hfill
    \begin{subfigure}[t]{0.49\textwidth}
        \centering
        \includegraphics[width=1\textwidth]{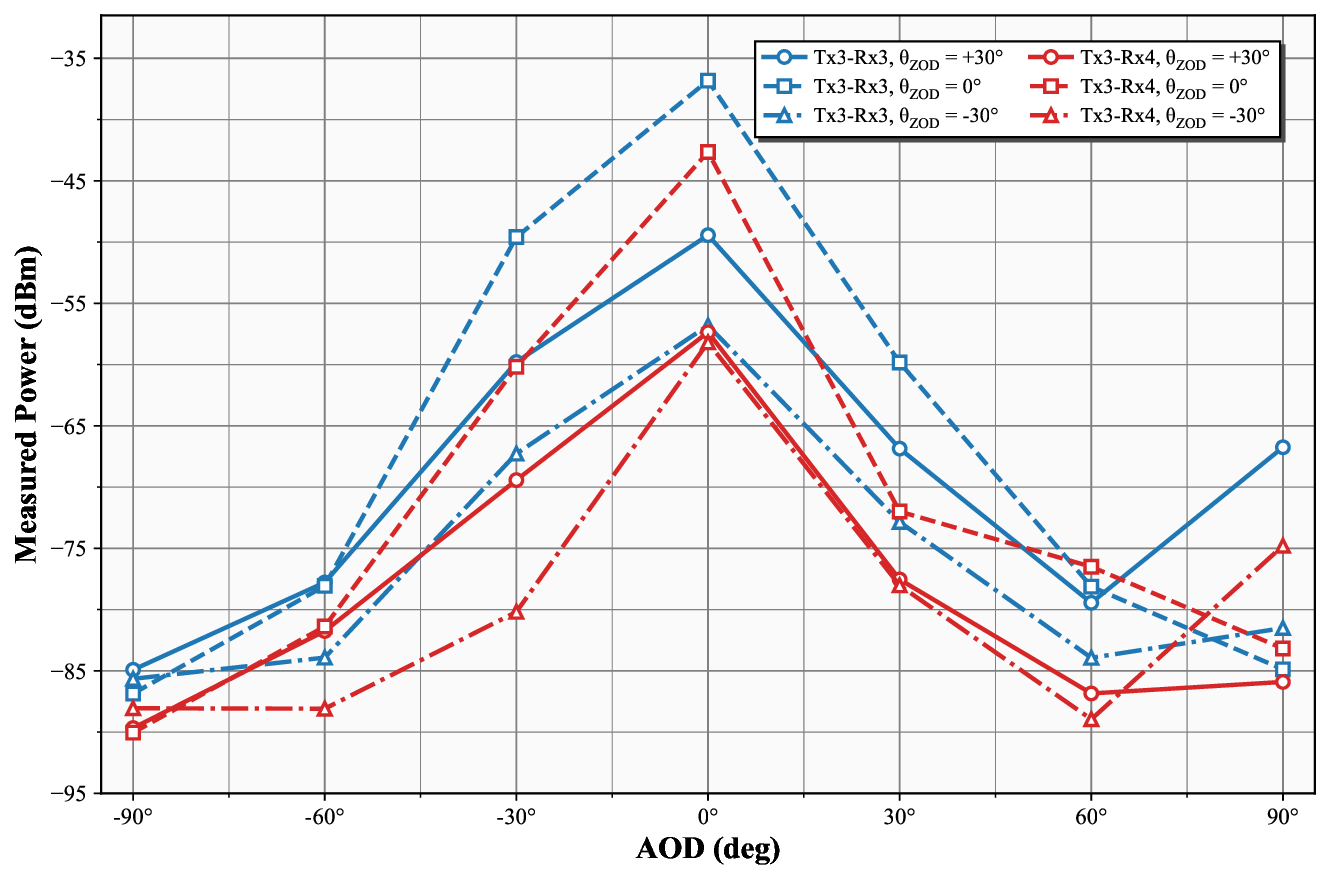}
        \caption{Tx3-Rx3/Rx4}       
    \end{subfigure}
    
    \vspace{0.3cm} 
    
    \begin{subfigure}[t]{0.49\textwidth}
        \centering
        \includegraphics[width=1\textwidth]{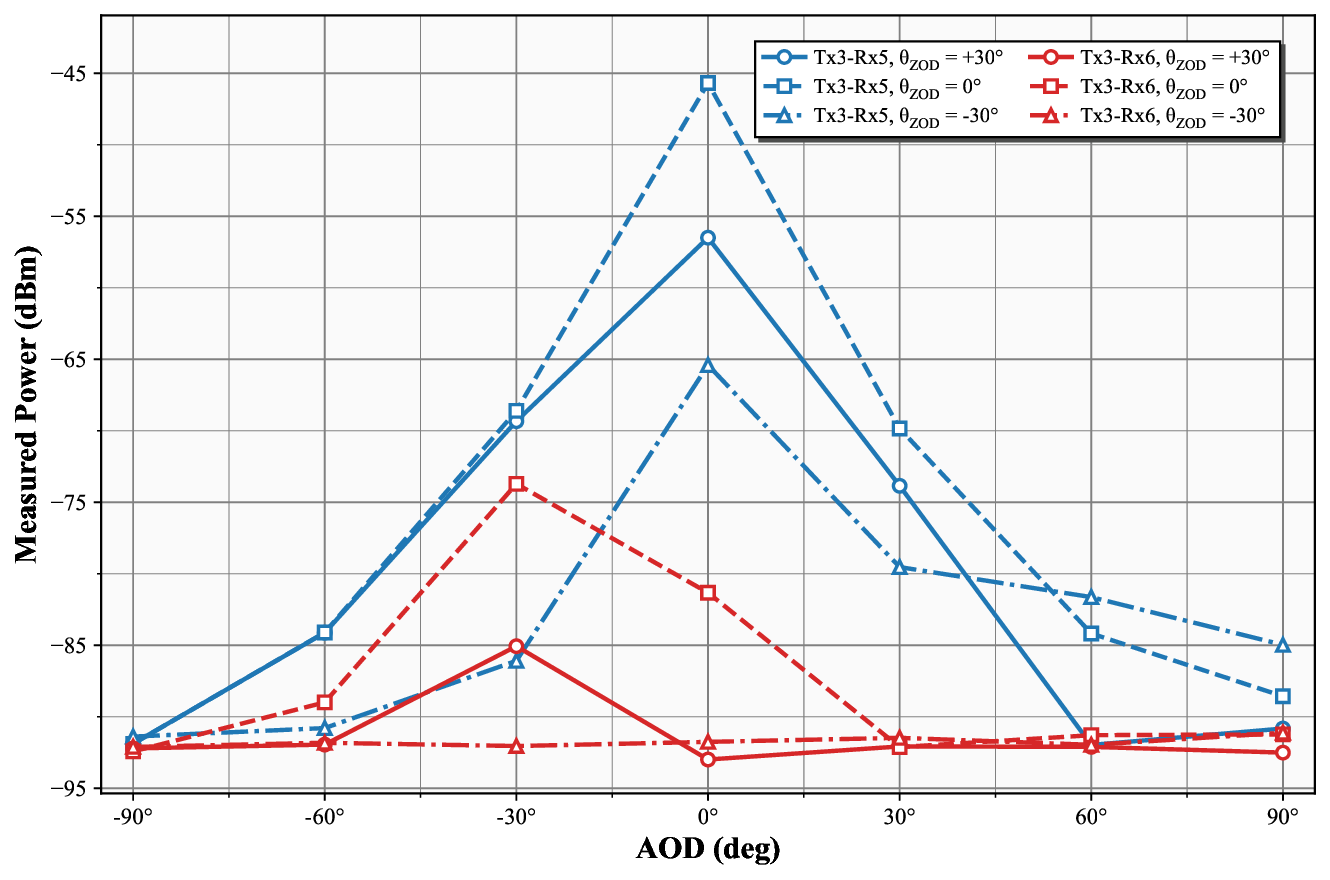}
        \caption{Tx3-Rx5/Rx6}        
    \end{subfigure}
    \hfill
    \begin{subfigure}[t]{0.49\textwidth}
        \centering
        \includegraphics[width=1\textwidth]{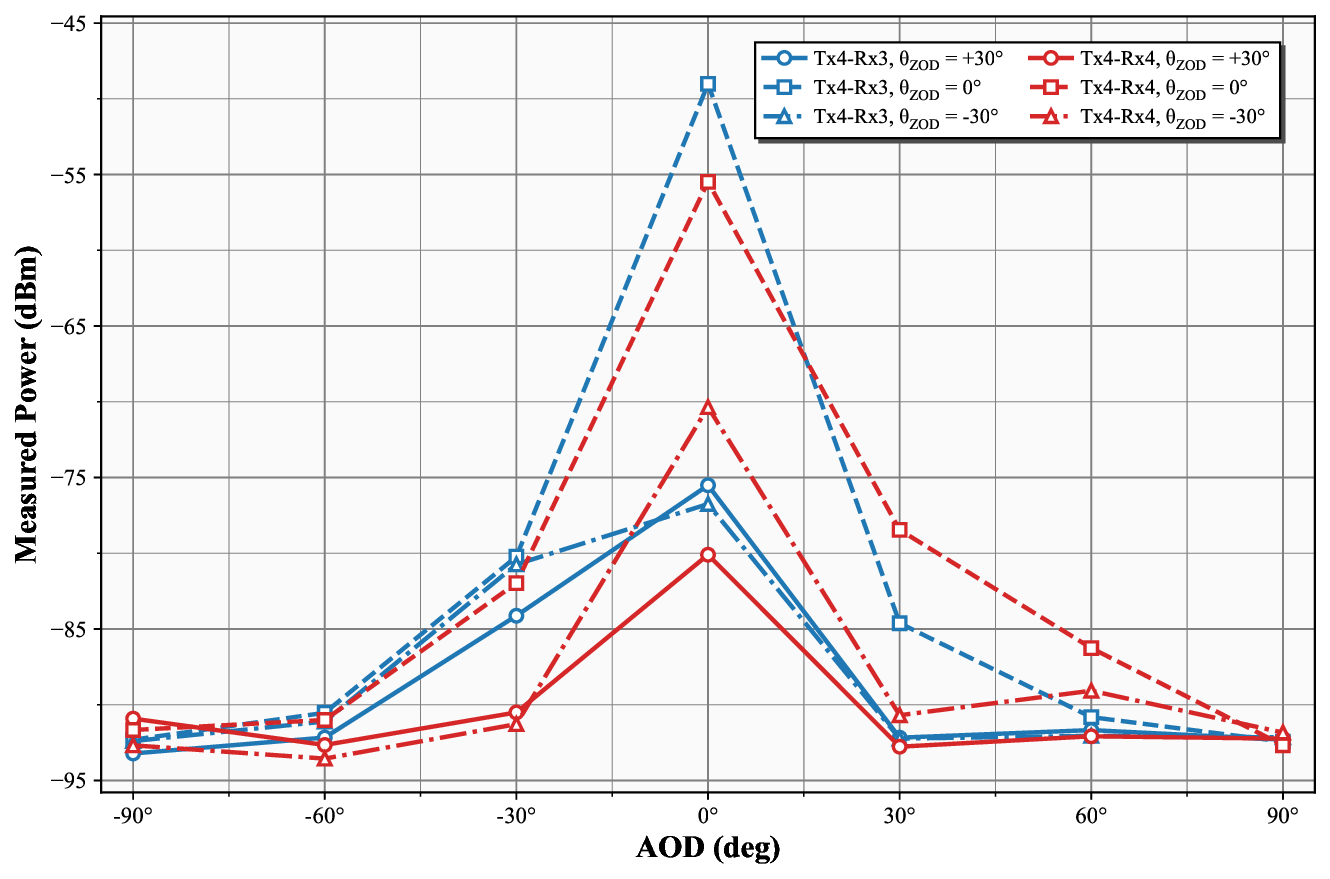}
        \caption{Tx4-Rx3/Rx4}  
    \end{subfigure}
    \caption{PAS on $(\phi_{AOD}, \theta_{ZOD})$ for different Tx-Rx Pairs.}
    \label{fig3}
\end{figure*}

In the free space, the electromagnetic wave propagation loss can be described using the free space path loss model, which is given by:
\begin{equation}
    \text{PL}^\text{FS}(f,d)[\text{dB}] = 20\log_{10}\left(\frac{4\pi fd}{c}\right),
\label{eq1}
\end{equation}
\noindent where $f$ denotes the carrier frequency, $d$ is the distance between the Tx and Rx, and $c$ is the speed of light. 
To accommodate the characteristics of different propagation scenarios, the 3GPP has provided reference path loss models for various environments. For the UMa scenario with antenna heights below 22.5 m, the 3GPP path loss model is expressed as \cite{3GPP38.901}:
\begin{equation}
    \text{PL}^\text{UMa}(f,d)[\text{dB}] = 28 + 22\log_{10}(d) + 20\log_{10}(f_c).
\label{eq2}
\end{equation}
As shown in Fig. \ref{fig2}, the path loss is recorded at Rx1 and Rx2 as the Tx position varies along the trajectory from Tx1 to Tx2. The Tx was positioned at 26 locations with 1-meter intervals along the connecting line between Tx1 and Tx2. The path loss at each position is calculated by:
\begin{equation}
    \text{PL}[\text{dB}] = P_{Tx} - P_{Rx} + G_{Tx} + G_{Rx} + G_{s},
\label{eq3}
\end{equation}
where $P_{Tx}$ is the transmitted power, $P_{Rx}$ is the received power, $G_{Tx}$ and $G_{Rx}$ are the Tx and Rx antenna gains, respectively, and $G_{s}$ represents the system gain including amplifiers, the received power at each position is selected by the maximum received power in the LoS alignment state.

Due to the geometric relationship, the distance between the Tx and Rx gradually decreases as the Tx moves from Tx1 to Tx2, which is reflected in Fig. \ref{fig2} as a gradual reduction in path loss. However, the measured path loss consistently exceeds both the free space path loss and the 3GPP model predictions, indicating the presence of additional influencing factors in the current measurement channel. To address this discrepancy, we consider a correction factor (CF) that has been validated in other UMa scenarios, which is defined in \cite{Wang17VTC}:
\begin{equation}
    \text{CF} = 1.0005 \times 10^{-4} h_{T}^2 - 0.0286 h_{T} + 10.5169,
\label{eq4}
\end{equation}
where $h_{T}$ is the Tx height in meters. Applying this correction factor to the 3GPP model yields the modified UMa path loss model:
\begin{equation}
    \text{PL}^\text{UMa}_\text{modified}[\text{dB}] = \text{PL}^\text{UMa} + \text{CF},
\label{eq5}
\end{equation}
To achieve better fitting accuracy and obtain a more suitable channel model for our measurement scenario, we also utilize the CI path loss model with a 1 m reference distance. This model has superior performance across diverse environments and frequency bands compared to conventional approaches \cite{Sun16TVT, Sun25TWC}. The CI model is formulated as follows:
\begin{align}
\text{PL}^\text{CI}(f, d)[\text{dB}] &= \text{PL}^\text{FS}(f, d_0) + 10n \log_{10}\left(\frac{d}{d_0}\right) + X_\sigma, \nonumber \\
&\quad \text{for } d \geq d_0, \text{ where } d_0 = 1 \text{ m,}
\label{eq6}
\end{align}
where $n$ is the PLE characterizing the decay rate, and $X_\sigma$ represents shadow fading modeled as a lognormal random variable with zero mean and standard deviation $\sigma$ in dB. The CI path loss model uses the $\text{PL}^\text{FS}$ at $d_0$ = 1 m as a fixed anchor point and fits the measured path loss data in log-distance scale, which is controlled by a single parameter n obtained via the minimum mean square error method. 

The fitting results presented in Fig. \ref{fig2} reveal distinct performance among the different path loss models. The height-dependent correction factor CF, defined by (\ref{eq4}), demonstrates over-correction when applied to the 3GPP path loss model at both Rx locations, leading to systematic deviations from the measured data. This observation suggests that the empirically derived correction factor, while validated in other UMa scenarios, may not adequately capture the unique propagation characteristics inherent to our low-altitude campus environment.
On the other hand, the CI model accurately fits the measured path loss variations across the entire Tx trajectory. The fitted PLEs of n = 2.28 for Rx1 and n = 2.31 for Rx2 are both marginally above the theoretical free space value of 2, confirming that LoS propagation constitutes the primary propagation mechanism at both Rx locations, despite the presence of environmental scatterers.
The shadow fading $\sigma$ reveals some spatial variations between the two received locations, with Rx1 ($\sigma=1.24$ dB) exhibiting a significantly larger standard deviation compared to Rx2 ($\sigma=0.75$ dB). This disparity can be attributed to the different propagation environments surrounding each Rx position. The shadow fading at Rx1 can be enhanced by multipath scattering from nearby buildings, vegetation, and other environmental obstacles.

\subsection{3D Spatial Power Angular Spectrum}

\begin{figure*}[!t]
    \centering
    \begin{subfigure}[t]{0.49\textwidth}
        \centering
        \includegraphics[width=1\textwidth]{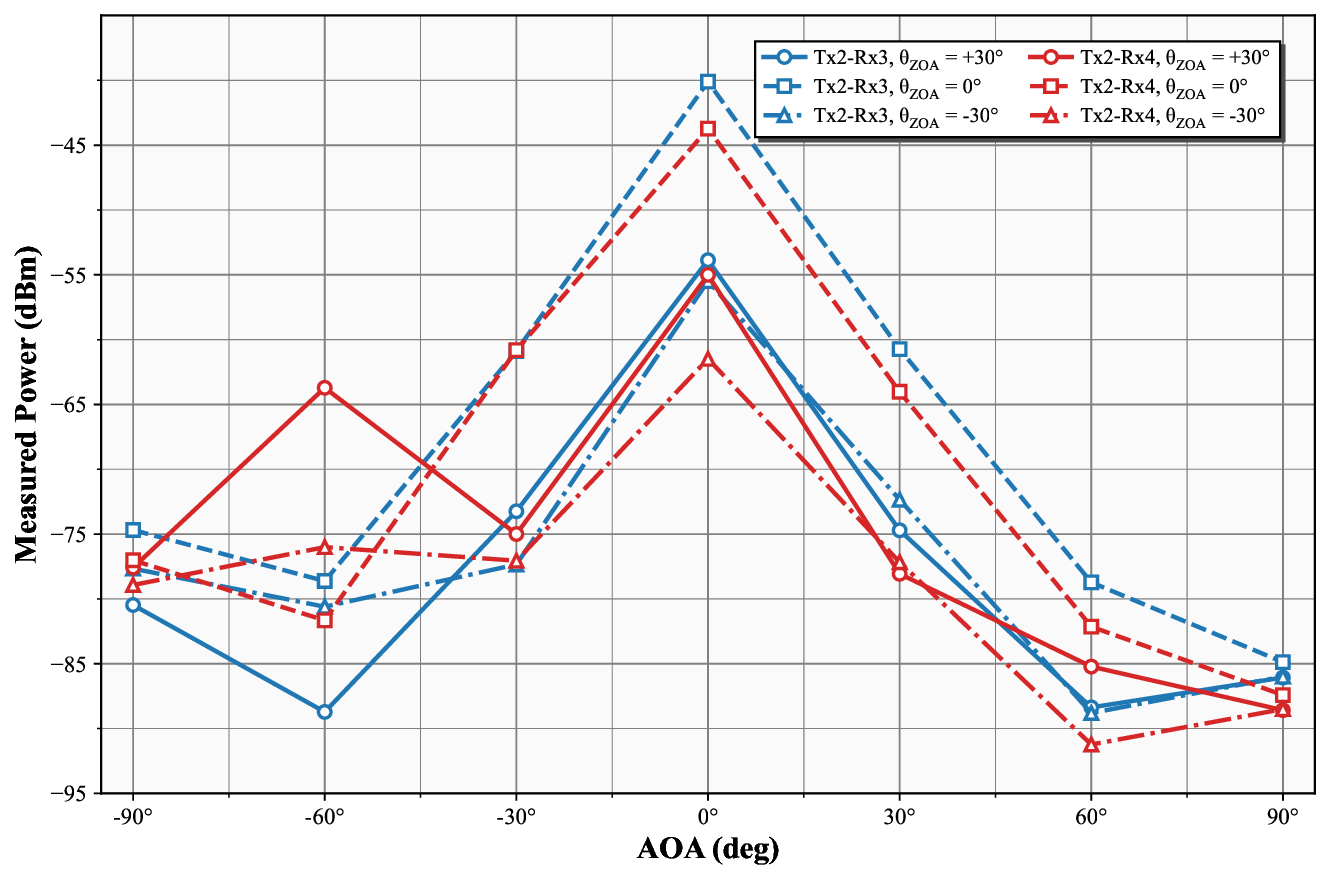}
        \caption{Tx2-Rx3/Rx4}        
    \end{subfigure}
    \hfill
    \begin{subfigure}[t]{0.49\textwidth}
        \centering
        \includegraphics[width=1\textwidth]{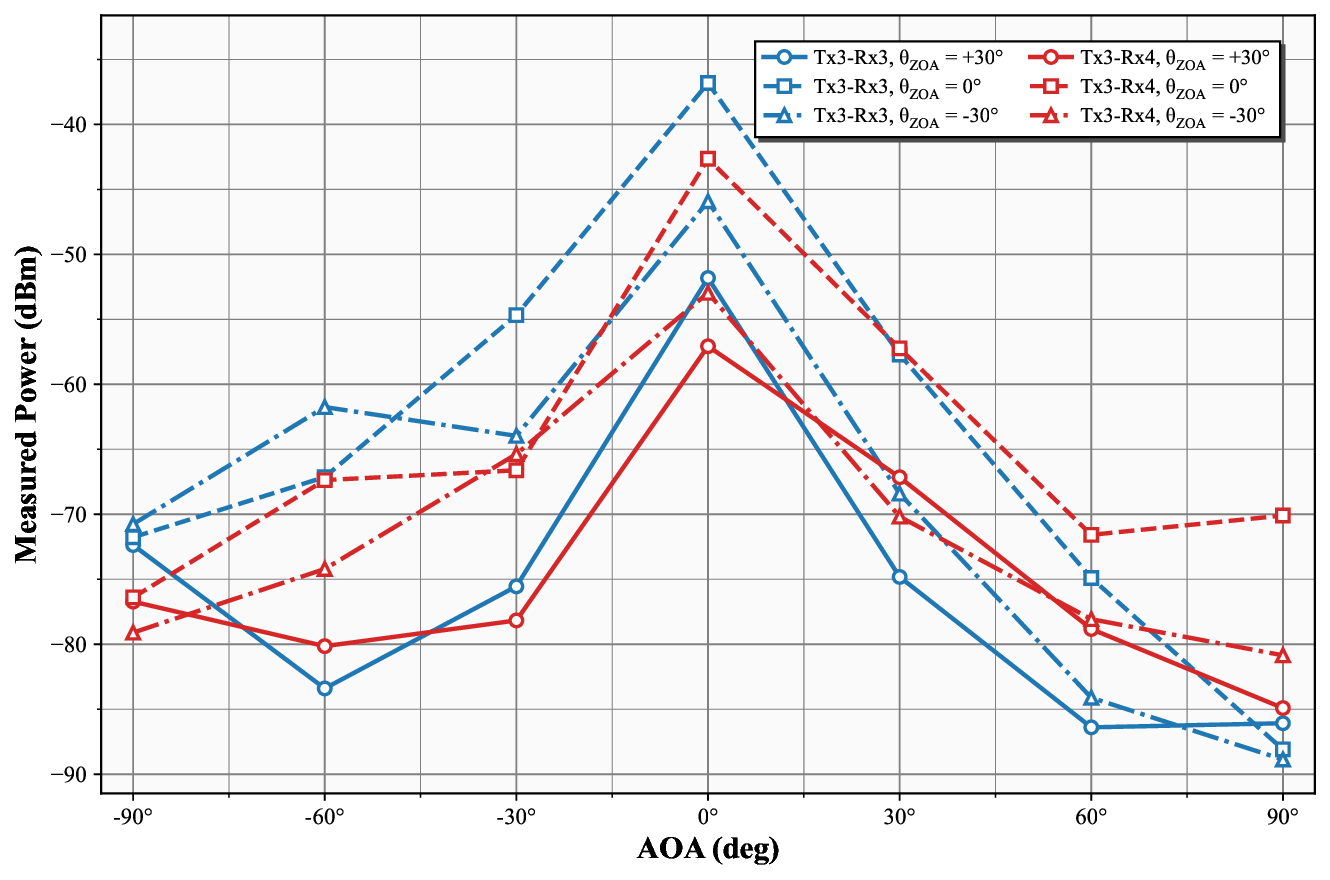}
        \caption{Tx3-Rx3/Rx4}       
    \end{subfigure}
    
    \vspace{0.3cm} 
    
    \begin{subfigure}[t]{0.49\textwidth}
        \centering
        \includegraphics[width=1\textwidth]{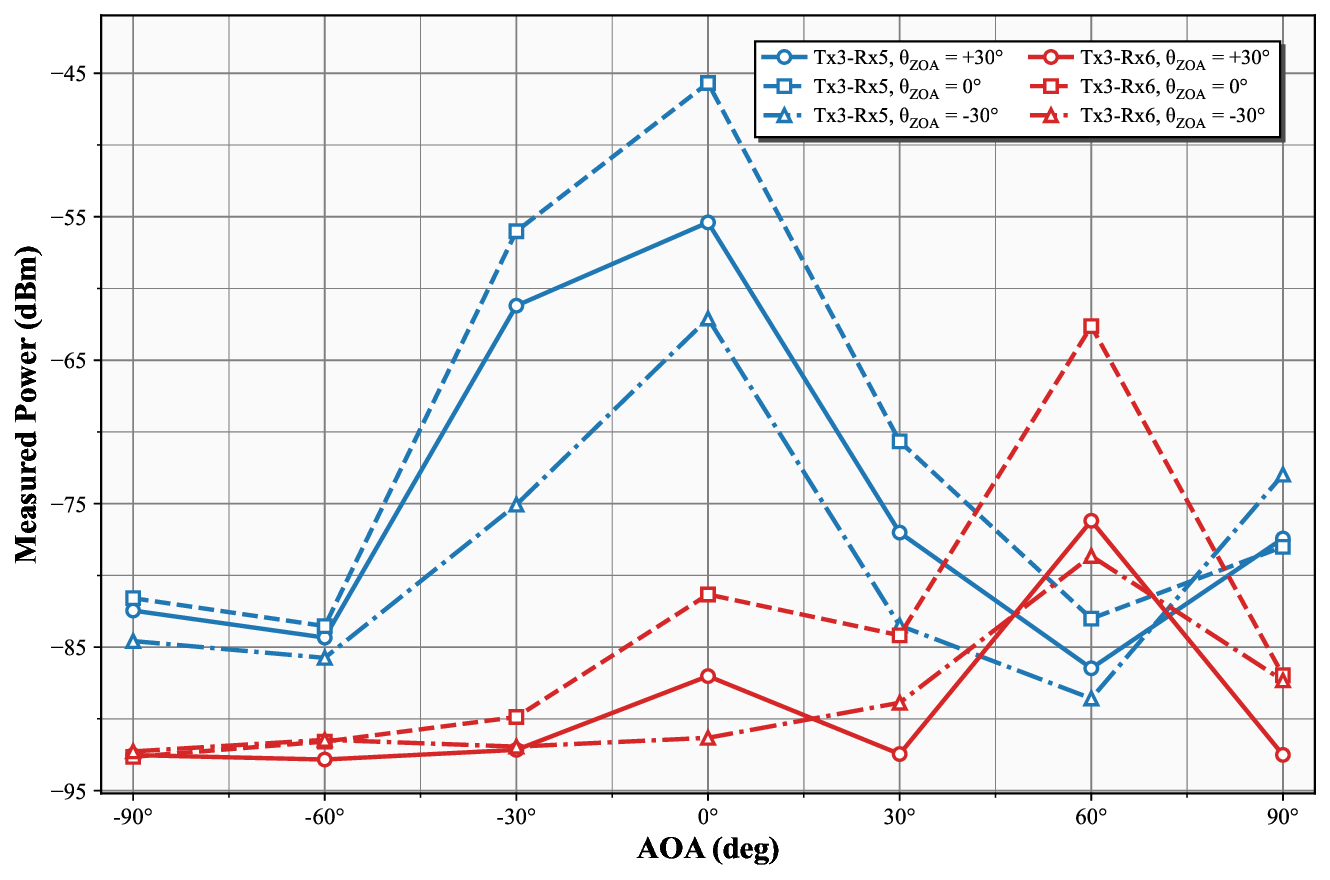}
        \caption{Tx3-Rx5/Rx6}        
    \end{subfigure}
    \hfill
    \begin{subfigure}[t]{0.49\textwidth}
        \centering
        \includegraphics[width=1\textwidth]{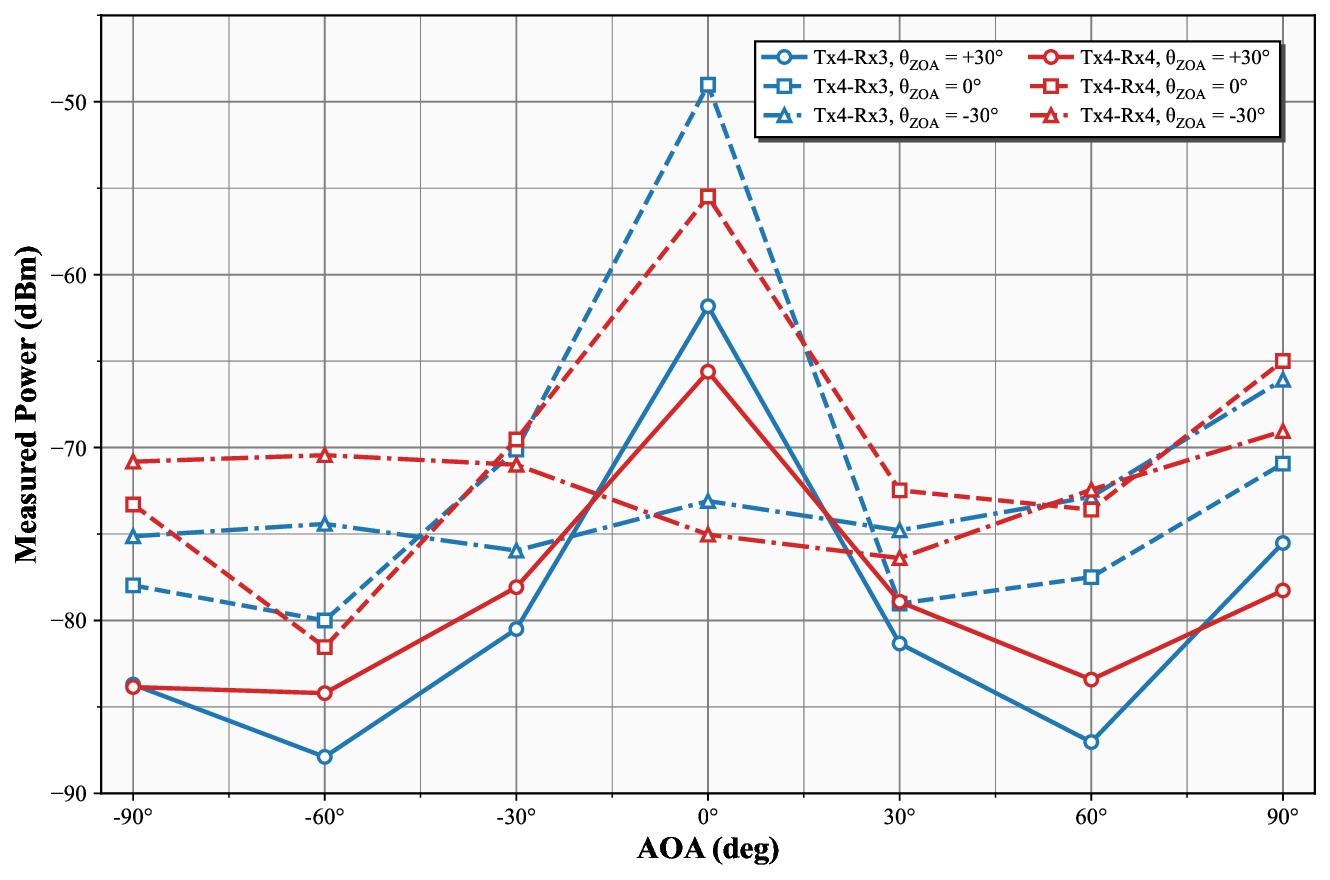}
        \caption{Tx4-Rx3/Rx4}  
    \end{subfigure}
    \caption{PAS on $(\phi_{AOA}, \theta_{ZOA})$ for different Tx-Rx Pairs.}
    \label{fig4}
\end{figure*}

Fig. \ref{fig3} and Fig. \ref{fig4} present the measured PAS on $(\phi_{AOD}, \theta_{ZOD})$ and $(\phi_{AOA}, \theta_{ZOA})$ for different Tx-Rx pairs, respectively. As shown in Fig. \ref{fig3}(a), due to the different distances from Rx3 and Rx4 to Tx2, on one hand, the closer Rx3 exhibits higher received power across the AOD range compared to the more distant Rx4. On the other hand, the shorter propagation distance makes the Rx3 more sensitive to ground reflections, as evidenced by higher received power when the Tx2 is tilted downward ($\theta_{ZOD} = -30$°) compared to upward tilting ($\theta_{ZOD} = +30$°). At the more distant Rx4, the received power levels for these two tilting configurations are comparable. This characteristic is further validated in Fig. \ref{fig3}(d), where for Tx4, Rx3 is located farther from the Tx4 than Rx4, resulting in the opposite behavior compared to the Tx2 scenario.

It should be noted that for the Tx4 transmission scenario, as shown in Fig. \ref{fig:measurement_setting}, there exists an obstruction along the LoS path between Tx4 and Rx4. This obstruction results in lower received power at the closer Rx4 across the AOD range, as observed in Fig. \ref{fig3}(d) and Fig. \ref{fig4}(d). Furthermore, in Fig. \ref{fig4}(d), when the Rx is tilted downward toward the ground ($\theta_{ZOA} = -30$°), the received power remains relatively uniform across different AOA directions, indicating weak contribution from LoS signal.
When the ground-level Rx is tilted downward, the Rx antenna height of 1.5 m creates geometric constraints that limit the reception of reflected signals from rooftop Tx due to unfavorable reflection angles. In contrast, the Rx maintains reasonable reception capability for signals from the ground-level Tx3, as demonstrated in Fig. \ref{fig4}(a) and Fig. \ref{fig4}(b).

For Rx5 and Rx6 positioned at the entrance and inside of the college corridor, respectively, Rx5 remains in a location visible to Tx3, while Rx6 is completely blocked within the corridor interior. In Fig. \ref{fig3}(c), Rx5 exhibits nearly symmetric received power distribution across the AOD range. In contrast, Fig. \ref{fig4}(c) shows significantly stronger received power on one side deviating from the AOA center, which can be attributed to the narrow spatial structure of the corridor entrance, causing uneven power distribution across different Rx pointing directions.
Interestingly, for Rx6, located completely inside the corridor, the AOD and AOA directions of the strongest received power points are toward the same side of the Tx3-Rx6 connection line. This indicates that the received signal at Rx6 primarily originates from environmental scattering contributions in specific directions, while the direct path signal suffers from severe shadowing due to the corridor structure.

\section{Conclusion}
In this paper, we presented a path loss model and analyzed 3D power-angular characteristics for the low-altitude channel based on extensive measurements at 10 GHz in an outdoor campus environment. In the measurement campaign, we simulated aerial measurements by moving the Tx along a predetermined trajectory at rooftop height, and measured the power angular spectrum by varying antenna pointing directions at both Tx and Rx sides. Path loss analysis revealed that the CI model can accurately capture the propagation behavior, providing reliable prediction capabilities for low-altitude channels. To further investigate the spatial richness of multipath propagation in low-altitude scenarios, comparative analysis of power angular spectrum characteristics was conducted across different Tx-Rx pairs. The observed channel conditions, which are significantly influenced by antenna height and ground scatterer coupling effects, indicate that future mid-band communication systems will require more precise and real-time channel prediction and modeling approaches.

\vspace{12pt}

\end{document}